# Cluster SIMS Microscope Mode Mass Spectrometry Imaging

András Kiss[1], Donald F. Smith[1], Julia H. Jungmann[1], Ron M.A. Heeren[1*]

[1] *FOM Institute AMOLF, Science Park 104, 1098 XG Amsterdam, The Netherlands*

*\* Author to whom correspondence should be addressed. Email: heeren@amolf.nl*

## Abstract

### Rationale

Microscope mode imaging for secondary ion mass spectrometry is a technique with the promise of simultaneous high spatial resolution and high speed imaging of biomolecules from complex surfaces. Technological developments such as new position-sensitive detectors, in combination with polyatomic primary ion sources, are required to exploit the full potential of microscope mode mass spectrometry imaging, i.e. to efficiently push the limits of ultra-high spatial resolution, sample throughput and sensitivity.

### Methods

In this work, a $C_{60}$ primary source is combined with a commercial mass microscope for microscope mode secondary ion mass spectrometry imaging. The detector setup is a pixelated detector from the Medipix/Timepix family with high-voltage post-acceleration capabilities. The system's mass spectral and imaging performance is tested with various benchmark samples and thin tissue sections.

### Results

We show that the high secondary ion yield (with respect to "traditional" monatomic primary ion sources) of the $C_{60}$ primary ion source and the increased sensitivity of the high voltage detector setup improve microscope mode secondary ion mass spectrometry imaging. The analysis time and the signal-to-noise ratio are improved compared to other microscope mode imaging systems, all at high spatial resolution.

### Conclusions

We have demonstrated the unique capabilities of a $C_{60}$ ion microscope with a Timepix detector for high spatial resolution microscope mode secondary ion mass spectrometry imaging.

**Introduction**

Mass spectrometry imaging (MSI)[1, 2] is a technique for the label free study and visualization of the distribution of multiple molecular species on complex surfaces, such as thin tissue sections. It has two main advantages over other common imaging techniques used. First it has chemical identification capabilities and second, no *a priori* knowledge of the sample is required. Matrix assisted laser desorption ionization (MALDI) has seen widespread use for mapping of intact biological molecules from complex surfaces. Spatial resolution in such experiments is typically limited to 10-50 µm, due to laser beam focusing. However, secondary ion mass spectrometry (SIMS) has a long history of sub-micrometer chemical imaging of a variety of sample substrates[3-9].

SIMS is the oldest ionization method used for mass spectrometry imaging[10]. Early primary ion sources for SIMS used atomic primary ions such as $Ga^+$, $Cs^+$ or $In^+$. These ion sources have the advantage that the ion beam can be focused to a very small spot size, thus a very high spatial resolution can be achieved (10s of nanometers)[11, 12] for MS imaging experiments. A drawback of these sources is the high degree of fragmentation of the secondary ions. The range of detectable ions is limited to elemental ions and small organic fragments such as $CH_3^+$ or $CN^-$, albeit at high spatial resolution. For many years, this restricted the use of SIMS to solid state physics and to the study of semiconductor surfaces. The introduction of polyatomic primary ion sources was one of the biggest advancements in the field[13-15]. It was shown by several groups that these sources have a higher secondary ion yield and they provide "softer" ionization[13, 16-20], thus opening the field of SIMS for biological research applications. However, these ion sources are more difficult to focus due to the strong space-charge effect associated with the combination of high primary ion current and small spot size. Thus, a compromise is needed between the primary ion current which is related to the secondary ion yield and spatial resolution. This is shown by the fact that the highest spatial resolution achieved with a $C_{60}$ primary ion source was 1 µm as opposed to the tens of nanometers with monoatomic primary ion beams. This was demonstrated by the Vickerman group which built a $C_{60}$ buncher Time-of-Flight (TOF) instrument for high spatial resolution imaging[19].

Typically, mass spectrometry imaging is performed in the microprobe mode, where a focused laser (MALDI) or primary ion beam (SIMS) measures the sample pixel by pixel. The spatial resolution is dependent on the spot size of the ion source in microprobe mode. *Microscope* mode imaging does not require highly focused beams. Rather, large beam sizes (usually around 200-300 µm) are used to desorb and ionize the molecules on the sample surface. After the desorption and ionization event, the spatial distribution of the ions is preserved as they travel through the mass spectrometer and then the ion image is projected onto a position-sensitive detector. The main advantage of this approach is that the spatial resolution is independent from the size of the laser focus or primary ion beam. In this way the analysis speed is significantly increased for the same area, spatial resolution and repetition rate compared to microprobe mode imaging. Additionally, it eases the difficulty of focusing the laser beam or primary ion source to very small spot sizes (few µm or less).

So far, the widespread use of microscope mode mass spectrometry imaging was limited by the lack of an appropriate position-sensitive detector with simultaneous measurement of time-of-flight and position information. The first detector used for microscope mode imaging consisted of the combination of a dual microchannel plate (MCP) with a phosphor screen and a charge-couple-device (CCD) camera[21, 22]. The main limitation of these detector systems is that they are not capable of recording the time-of-flight of the arriving ions, only their position. Thus, the selection of an ion of interest with an electrostatic blanker in the mass spectrometer is needed to record a selected-ion image. Because of the sample damage due to the ionization process, this is only possible for a few ions before the sample is depleted. It is also very time consuming because the sample needs to be imaged separately for every ion of interest.

Recently, other detectors, such as delay-line detectors, have been tested for microscope mode imaging[23, 24]. These detectors have the capability to record the time-of-flight and the spatial position of an ion simultaneously. However, they lack multi-hit capabilities for typical mass spectrometry imaging event rates. As a result, these detection systems require low ion loads and are only well suited for SIMS experiments and not MALDI. Additionally, there is no direct feedback during measurement because of the time consuming image reconstruction process which makes optimization of measurement parameters difficult.

The latest development in the field of microscope mode mass spectrometry imaging has been the introduction of pixelated detectors, such as the Medipix/Timepix detector family[25-30]. In these detectors, every pixel acts as an individual detector capable of recording the time-of-flight of the arriving ions with respect to an external trigger signal. Combined with MCPs, they offer multiplexed ion detection capabilities where every ion hit is registered by multiple pixels. This multiplexed detection results in increased sensitivity for these systems. Additionally, the spatial information is determined by the pixel address of each pixel. The capabilities of the Timepix detectors were previously demonstrated for microscope mode MALDI imaging[27-30].

In earlier work, we presented the first example of microscope mode SIMS imaging with a Timepix detector[31]. The potential of this detection system for SIMS microscope mode MSI, namely superior spatial resolving power and signal-to-noise was demonstrated. Also, these first experiments revealed that this initial setup lacked the sensitivity and speed needed for practical use in biological studies. Limitations were the low secondary ion yield of the gold primary ion source, the lack of ion post-acceleration in the Timepix setup and the low repetition rate of the Timepix detector. Also, negative mode measurements were not possible with this setup.

This work presents an improved Timepix detector setup[29] with ion post-acceleration and negative ion detection capabilities in combination with a $C_{60}$ primary ion source. This combination offers improvements in most of the crucial parts of the system such as sensitivity, speed and mass range. In particular, the use of a polyatomic $C_{60}$ primary ion source promises improved ion yield performance as compared to the gold liquid metal ion source and is evaluated in detail in this work. The capabilities of the system for microscope mode $C_{60}$ SIMS imaging were demonstrated with various benchmarks and thin tissue sections.

**Materials and methods**

**The Medipix/Timepix detectors**

The Timepix detector used for this project is the member of the Medipix/Timepix detector family[32-35]. These are active pixels detectors developed by the Medipix collaboration at the European Organization for Nuclear Research (CERN, Geneva, Switzerland). They are based on Complementary Metal-Oxide Semiconductor (CMOS) technology[36]. The Timepix application-specific integrated circuit (ASIC) is an improved version of the Medipix ASIC with two new measurement modes in addition to the simple particle counting mode. One is the Time-over-Threshold mode (TOT mode) in which every pixel registers how long it is above a certain charge threshold level. The other operating mode, and the one used for mass spectrometry, is the Time-of-Flight mode (TOF mode). One Timepix chip consists of 256×256 pixels, each capable of individually measuring the time-of-flight of ions hitting the detector with respect to an externally applied trigger signal. Each clock cycle is 10 ns wide and the maximum measurement interval of the Timepix detector is 118 µs. The size of an individual pixel is 55 µm×55 µm. It is also possible to build 2×2n arrays of Timepix chips if bigger detectors are necessary for the experiments.

**Microscope mode mass spectrometer**

A TRIple Focusing Time-of-Flight (TRIFT II) mass spectrometer (Physical Electronics, Inc., Chanhassen, MN, USA) is used. The mass spectrometer is equipped with a 20 keV $C_{60}$ primary ion source (Ionoptika, Chandlers Ford, Hampshire, United Kingdom) and a high voltage Timepix detector setup[29]. The primary ion source is operated at 20 keV primary ion energies with the $C_{60}^{2+}$ ion selected for high primary ion beam current. The primary ion beam uses a pulse length of 60 ns and is then bunched for better spectral resolution. All the measurements are done in static SIMS mode, where the primary ion dose is well below the static limit ($1 \times 10^{13}$ ions/cm$^2$).

The high voltage Timepix setup installed on the instrument has been described in details elsewhere[29]. Briefly, it consists of a 2×2 array of Timepix chips behind a chevron MCP stack. The MCPs are operated at a bias of 1.375 kV for the positive mode and 1.4 kV for the negative mode experiments, unless it is stated otherwise. The data readout system uses the ReLAXD (high Resolution Large Area X-ray Detector) readout board which has a readout speed of 1 Gbit/s[37, 38] and is operated at a frame rate of 10 frames/s. The chips are cooled with a Peltier element based active cooling system. The entire detector setup can be floated at +12 kV (for negative ion mode) or -8 kV (for positive ion mode) with the use of the TRIFT II mass spectrometer's high voltage power supply. This offers additional post-acceleration of the secondary ions before they reach the MCPs. Both the Timepix/RelaXD system and the $C_{60}$ source are triggered from the TRIFT II mass spectrometer master trigger, which is down sampled 100 times to 10 Hz. For the data acquisition the Pixelman data acquisition software is used[39].

The Timepix detector is operated in TOF mode. In all modes, a typical, sparse data frame contains the x- and y-coordinate of every triggered pixel. Additionally, in TOF mode, the data file lists the time-of-flight. In the TOF mode, every measurement frame contains the TOF information obtained from a single primary ion pulse. The mass spectrum is reconstructed by making a histogram of the TOF values from the separate frames. Standards are used to calculate mass calibration parameters, which are then applied to the TOF spectra. Total ion images are constructed by summing all of the individual frames. Selected ion images are plotted by extracting the pixel positions and intensities for a selected mass spectral peak.

**Samples**

Various benchmark samples are used to test the imaging and mass spectral performance of the system. These include brilliant green dye

(green Staedtler Lumocolor 318-5 permanent marker, Staedtler Mars GmbH & Co. KG, Nuernberg, Germany) as well as a 1 mg/mL solution of a mixture of different chain length polyethyleneglycols (PEG 200-3500) mixed with 7 mg/mL α-cyano-4-hydroxycinnamic acid (CHCA) in 1:1 ratio (1 µL deposited). The benchmark samples are placed on an Indium-Tin oxide (ITO) coated glass slide (4-8 Ώ resistance, Delta Technologies, Stillwater, MN, USA). A hexagonal transmission electron microscopy (TEM) grid (700 mesh, G2760N, 3.05 mm diameter, 37 µm pitch, 8 µm bar width; Agar Scientific Limited, Stansted, United Kingdom) is placed on the top of the samples. For biological tissue imaging, 12 um thick coronal mouse brain (male balb/c mouse; Harlan Laboratories, Boxmeer, The Netherlands) sections are used. The brain is sectioned in a Microm HM525 cryomicrotome (Thermo Fisher Scientific, Walldorf, Germany) and the sections are placed on an ITO coated glass slide. The sections are kept at -20 $^{o}$C until further use. Before measurement the samples are dried in a vacuum desiccator and are measured without any further sample preparations steps.

**Results and discussion**

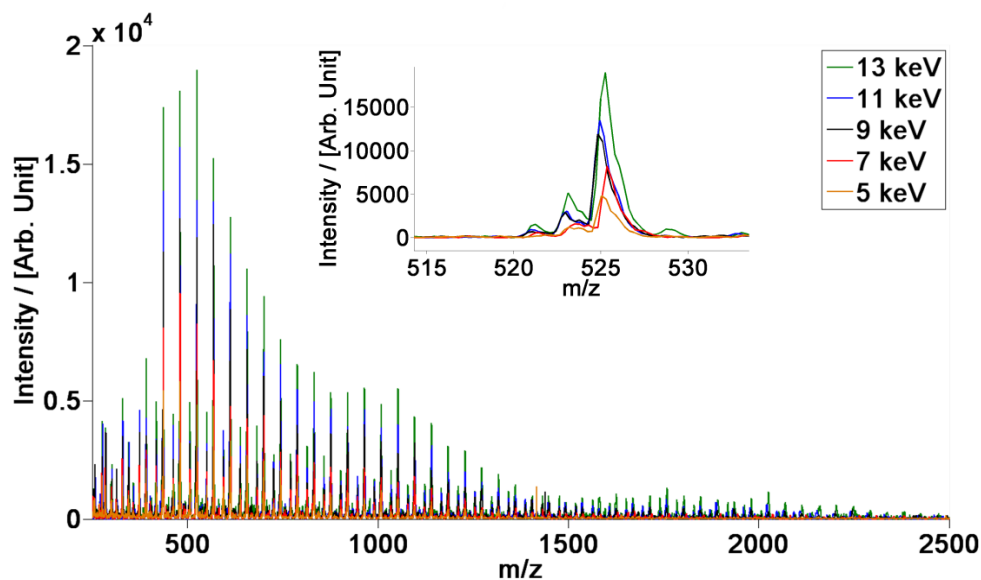

**Figure 1. Mass spectra of a mixture of polyethylene glycol measured with $C_{60}$ SIMS on the Timepix detector at different ion acceleration energies. The inset shows a selected PEG peak at *m/z* 525. Higher ion acceleration energies improve the signal-to-noise ratio and extend the accessible upper mass range.**

As a first experiment the system's spectral quality is assessed. Also, the effect of the increased post-acceleration of the ions on the mass spectra is systematically studied. A mixture of PEGs is used and 10,000 frames at different total ion acceleration energy values between 5 keV and 13 keV. The MCP is operated at an MCP bias of 1.5 kV. Figure 1. shows an overlay of the PEG mass spectra obtained at the different total ion acceleration energies. The sodiated molecular ions of the PEGs are detected in this measurement. Increasing the acceleration energy of the secondary ions has a two-fold effect on the mass spectra. First, the intensity of the detected ions increases. Thus, the higher post acceleration yields a better signal-to-noise ratio (S/N of 184 for the ion at *m/z* 525 at 13 keV total acceleration energy and S/N 55 at 5 keV total acceleration energy). Also, the accessible mass range is significantly increased at the highest energy values. This is due to the higher ion energies, where more ions have a chance to start an electron cascade in the MCP. Also, higher mass ions that would not have the necessary energy to start an electron cascade can be detected due to the additional post-acceleration capabilities of the system. It is important to note that the same sample measured for 10,000 frames with an Au primary ion source with a higher MCP gain results in a spectrum with a S/N of 37, which is five times lower than the S/N achieved with the combination of the $C_{60}$ source and the high voltage Timepix setup, and mass range between *m/z* 0 and 1200 (see supplementary figure S1). Thus, the use of the $C_{60}$ primary ion source results in a significant increase in spectral quality, even at low post-acceleration, due to the higher secondary ion yields associated with such polyatomic primary ion sources.

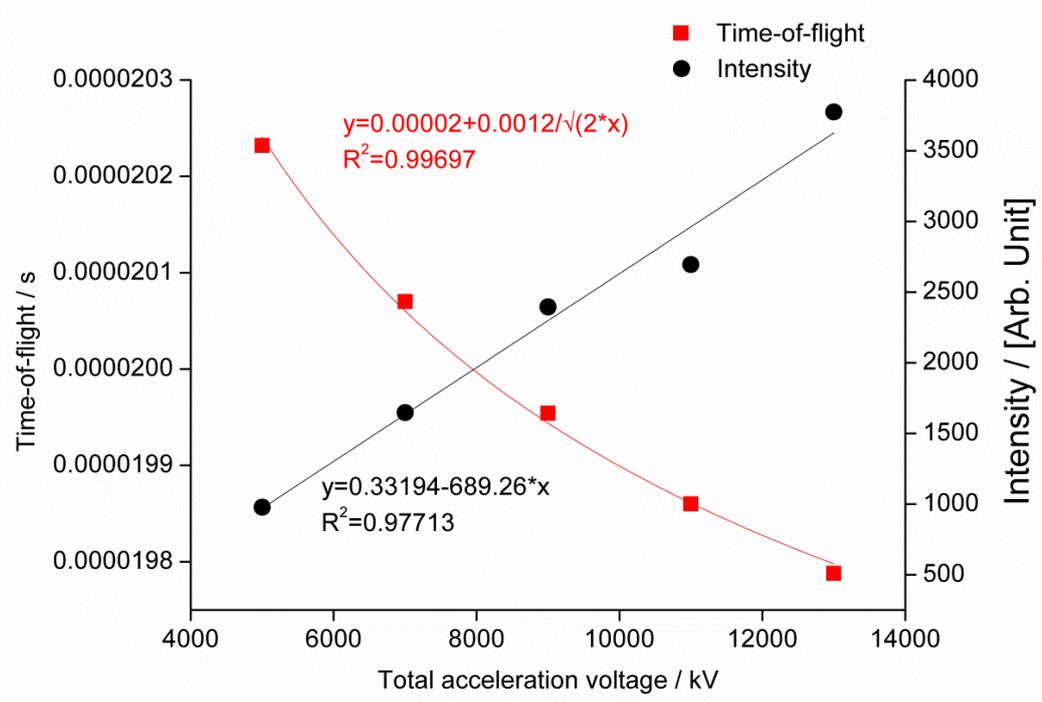

**Figure 2. Intensity (black) and the time-of-flight as the number of clock cycles (red) of the PEG ion at *m/z* 526 as a function of the ion acceleration energy.**

The high voltage setup has two main effects on the spectral quality. The first one is the increase of the secondary ion intensities discussed in the previous paragraph. The second effect is a decreasing time of flight of the same ion as a function of the acceleration voltage. To demonstrate these effects the measurements shown in Fig. 1 are divided into 5 x 2,000 frame segments. Figure 2. shows the average intensities and average time-of-flight values of these five smaller datasets as a function of the ion acceleration energies. The time-of-flight of the selected ion changes as a quadratic function of the acceleration voltage. This is in agreement with the TOF analyzers calibration equation (1)

$$t = \frac{d}{\sqrt{2U}} \sqrt{\frac{m}{q}} \qquad (1)$$

where t is the time-of-flight of the ion, d is the length of the ion's flight path, m is the mass of the ion, q is the charge of the ion and U is the acceleration voltage. Analysis of the dependence of ion intensity demonstrates a linear increase with ion kinetic energy, as expected.

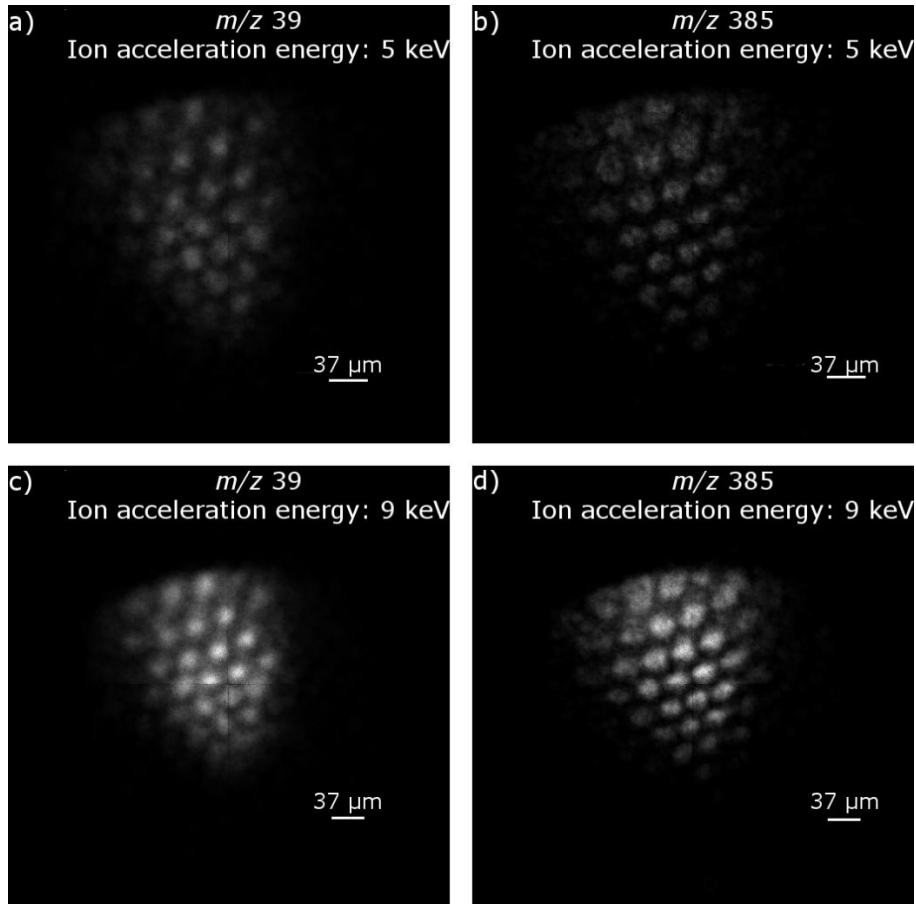

**Figure 3. Comparison of the image quality at different ion acceleration energies. Selected-ion images of potassium at *m/z* 39 (a, c) and brilliant green dye at *m/z* 385 (b, d) are plotted on the same intensity scale. The detector is operated at total ion acceleration energies of 5 keV (a, b) or 9 keV (c, d).**

Figure 3 demonstrates the imaging capabilities of the system on brilliant green dye underneath a TEM grid, in the positive ion mode. The same area of the sample is measured with the detector at ground (Figure 3a and 3b) and at 4 kV (Figure 3c and 3d). The selected-ion images for the comparison are the cation of the brilliant green dye (*m/z* 385) and potassium ion (*m/z* 39) which both localize in the holes of the grid. The size of the primary ion beam is ~224 µm × 272 µm and the images are reconstructed from 20,000 frames. Fig. 3 shows that the high voltage setup is capable to provide the same image quality as the previous detector setup (where the detector is held at ground potential). The effects of the higher ion acceleration on the image quality are a better image contrast, due to the higher S/N ratio, and a smaller magnification factor, due to the shorter time that the ions spend in the magnification region of the instrument. The observed spatial resolving power is ~7 µm (see supplementary figure S2.) with a pixel size of 900 nm. This value is close to the previously reported[31] spatial resolutions for the Timepix setup without high voltage capabilities and for a microprobe mode reference measurement system that uses the combination of an MCP with a time-to-digital converter (TDC), both on a TRIFT mass spectrometer.

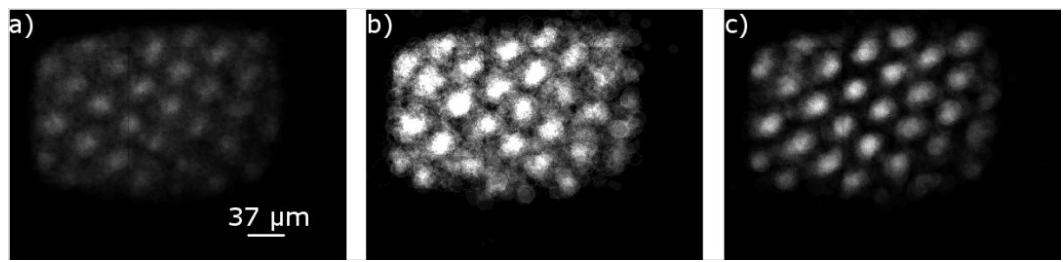

**Figure 4. Negative mode imaging of brilliant green dye under a TEM grid. Selected-ion images show the distribution of $C_2H^-$ at *m/z* 25 (a), $C_4H^-$ at *m/z* 49 (b) and $HSO_4^-$ at *m/z* 97 (c).**

The previous Timepix setup was only operable in positive ion mode. One of the advantages of the new high voltage setup is the capability to detect negative ions as well. Thus, it is capable of providing complementary information to the positive mode measurements. This new functionality of the system was previously demonstrated on peptide and protein standards with MALDI ionization[29]. However, no negative mode images were recorded thus far with a pixelated detector. Figure 4. shows the first example of negative mode microscope mode SIMS imaging with a Timepix detector. The sample is the same grid standard used for the positive mode measurements shown in Fig. 3. The detector is operated at an ion acceleration energy of 12.5 keV. As expected, the ions detected are mostly the standard small negative ions common to SIMS, such as $[C_2H]^-$ at *m/z* 25, $(C_4H)^-$ at *m/z* 49; or $(PO_3)^-$ at *m/z* 79. These ions are abundant in the holes of the sample which means they are present in the green dye underneath the grid. The image quality is similar to the positive mode image quality as can be seen in supplementary Figure S2. This evaluation reveals a resolving power of 7.5 µm in negative mode which is in accordance with the images acquired in positive ion mode.

The fivefold increase in sensitivity achieved with the new setup makes the imaging of biological samples possible. Figure 5 exemplifies the system's improvement for the imaging of biological samples. This first example of a mosaic microscope mode SIMS imaging with a pixelated detector shows complementary positive and negative ion mode images of a biological tissue sample. In particular, the sample is half of the anterior commissure area of a mouse brain section as can be seen on the hematoxylin and eosin (H&E) stained image (Figure 5a). The images are reconstructed from 4 tiles. The step size between each tile is 100 µm, such that there is sufficient overlap between the tiles to correct for the elliptic shape of the primary ion beam spot during the reconstruction of the image. In both positive and negative mode, 20,000 frames per tile are collected. One measurement takes 80,000 frames compared to the single tile imaging with the gold primary ion source that took more than 300,000 frames to achieve similar image quality. This means that as a result of the improvements, ~80% fewer frames are needed to measure an area that is roughly four times the size as the previously reported microscope mode SIMS image with the Timepix detector. Further, the increased secondary ion yield of the $C_{60}$ source reduces the measurement time to only half of the time reported earlier for the single tile microscope mode SIMS measurement. Also, intact phospholipids are detected from the tissue (see supplementary figure S3).

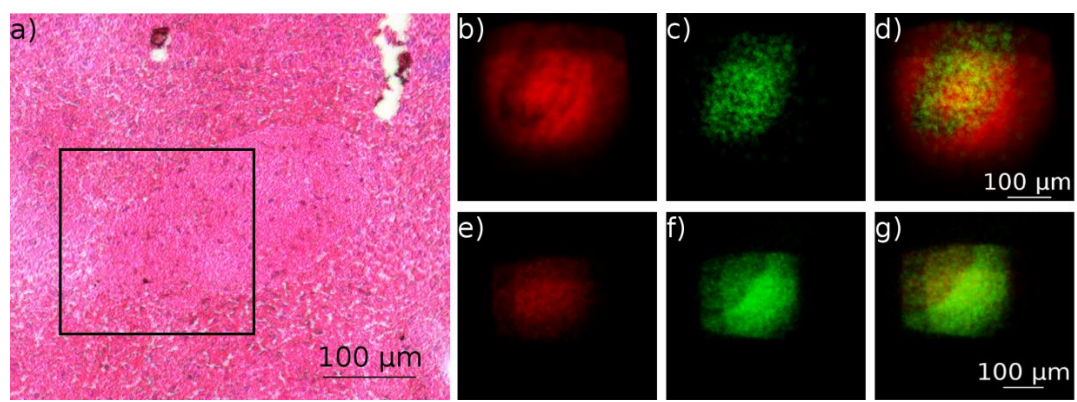

**Figure 5. Microscope image of a H&E stained mouse brain (a), positive (b, c, d) and negative (e, f, g) microscope mode SIMS image of the anterior commissure region of the mouse brain. The selected ion images show the distribution of sodium at *m/z* 23 (b), the cholesterol fragment [M+H-H$_2$O]$^+$ at *m/z* 369 (c), CH$^-$ at *m/z* 13 (e) and PO$_3^-$ at *m/z* 79 (f) and the overlay of the two positive (d) and negative mode (g) images**

It is possible to distinguish the anterior commissure in both positive and negative mode. The positive mode selected ion images show the distribution of the sodium ion at *m/z* 23, which has higher abundance in the tissue surrounding the anterior commissure and a cholesterol fragment ([M+H-H$_2$O]$^+$) at *m/z* 369, which localizes in the anterior commissure. The ions selected for the negative mode images are (CH)$^-$ at *m/z* 13 with a homogenous distribution in the imaged area of the tissue and (PO$_3$)$^-$ at *m/z* 79, which is localized in the tissue around the anterior commissure.

**Conclusions**

For the first time, the combination of a $C_{60}$ primary ion source and a pixelated detector system with ion post acceleration capabilities was used for high spatial resolution microscope mode SIMS imaging in both positive and negative ion mode. The combination of the higher secondary ion yield of the $C_{60}$ primary ion source and the higher ion post acceleration possibilities of the new Timepix setup resulted in an increased signal-to-noise ratio and wider mass range compared to the earlier microscope mode SIMS experiments. The acquisition time necessary to achieve high quality data was also significantly reduced. The image quality, namely the spatial resolving power, is comparable to the earlier Timepix based microscope mode SIMS studies. It is also possible, for the first time, to measure microscope mode SIMS images in negative mode with a pixelated detector.

Timepix based microscope mode SIMS imaging opens up new possibilities, such as the fast 3D imaging of complex samples with sub-cellular spatial resolution. However, several challenges such as the speed of the readout system and the single stop nature of the pixels remain that require further improvements on the detector level. These include a faster, 1 kHz readout system, multi-hit capabilities and a new, compact data format. Also, online, on-the-fly data analysis capability integrated on the chips or on-board the readout electronics would be advantageous for future applications and are subject to present and future studies. Microscope mode SIMS imaging with a pixelated detector shows promising results for high resolution mass spectrometry imaging of biological systems.

**ACKNOWLEDGEMENTS**

This work is part of the research program of the Foundation for Fundamental Research on Matter (FOM), which is part of the Netherlands Organisation for Scientific Research (NWO). Part of this research is supported by the Dutch Technology Foundation STW, which is the applied science division of NWO, and the Technology Program of the Ministry of Economic Affairs, Project OTP 11956. This publication was supported by the Dutch national program COMMIT and the Netherlands Proteomics Center. The authors acknowledge Sheng Chen (Physical Electronics) for excellent support during this project by sharing his knowledge on the TRIFT II system and Paul Blenkinsopp and Rowland Hill (Ionoptika) for their help with $C_{60}$ primary ion source. The authors are thankful to Berta Cillero Pastor who performed the pathological staining of the tissue samples, Frans Giskes and Ronald Buijs for their technical support and Ivo Klinkert for his support with the data processing.

# Cluster SIMS Microscope Mode Mass Spectrometry Imaging

András Kiss[1], Donald F. Smith[1], Julia H. Jungmann[1], Ron M.A. Heeren[1*]

[1] *FOM Institute AMOLF, Science Park 104, 1098 XG Amsterdam, The Netherlands*

*\* Author to whom correspondence should be addressed. Email: heeren@amolf.nl*

# SUPPLEMENTARY FIGURES

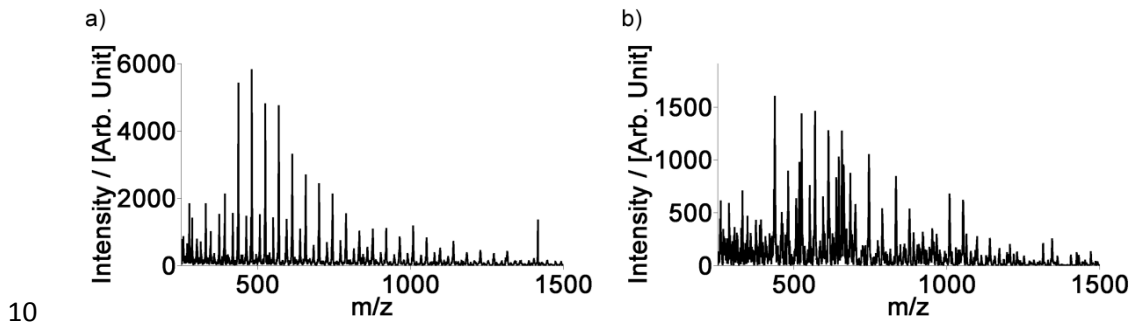

Supplementary figure S1. Comparison of a PEG spectra recorded with the $C_{60}^+$ Timepix setup (a) and with the $Au^+$ LMIG Timepix setup (b)

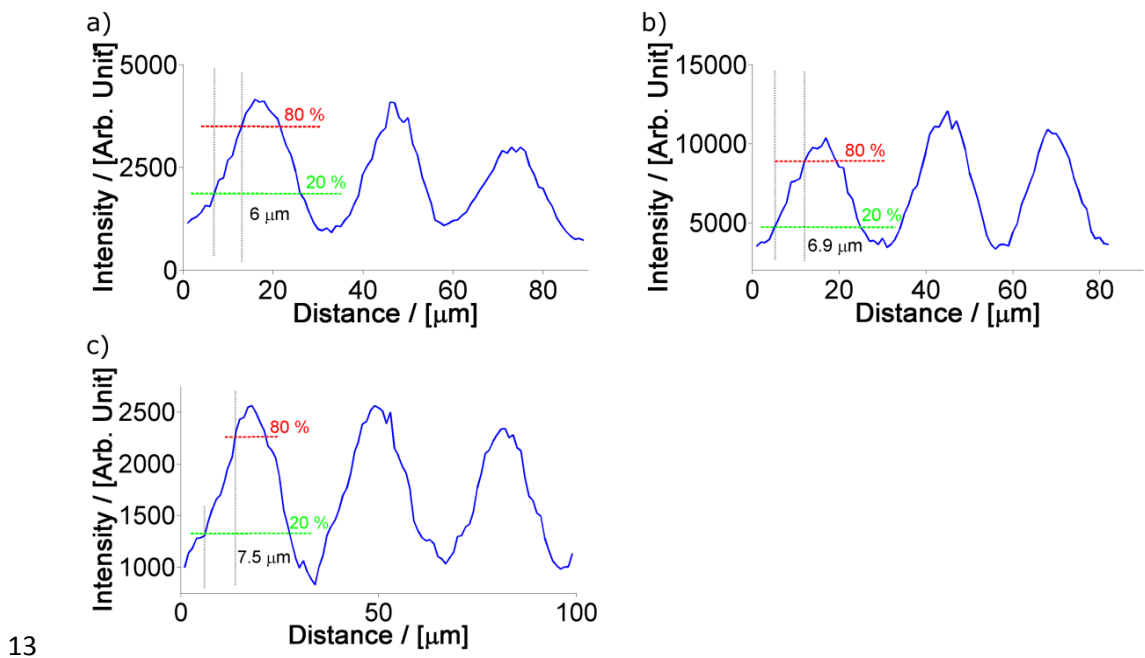

Supplementary figure S2. Line scans from the total ion image of the TEM grid in positive mode at a total acceleration energy of 5 keV (a) and 9 keV (b) and in negative mode at the total acceleration energy of 12.5 keV (c)

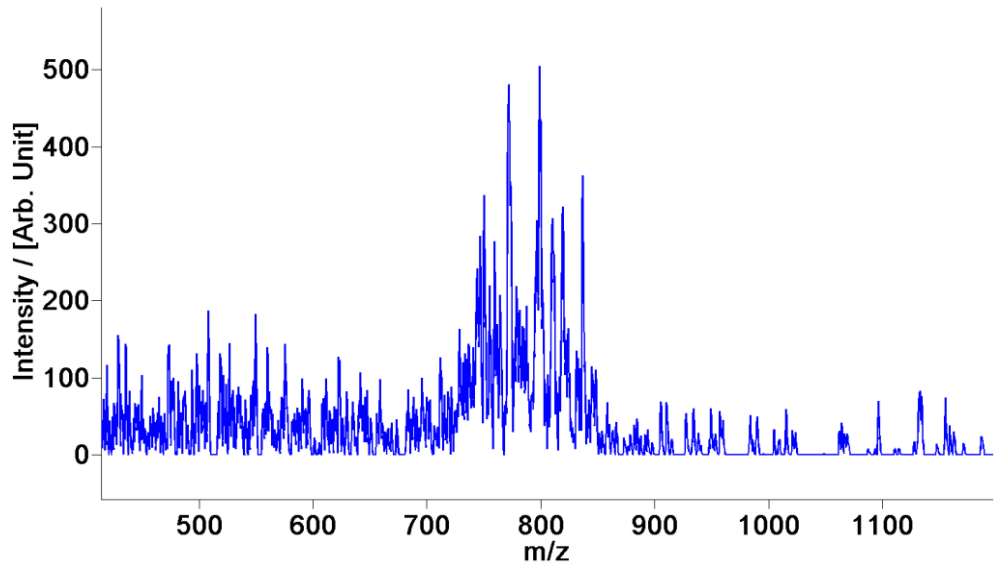

**Supplementary figure S3. Mass spectrum from the positive mode tissue imaging experiment. The spectrum shows the detected intact phospholipids from the tissue**